\documentclass[aps,twocolumn,showpacs]{revtex4}
\usepackage{graphicx}
\usepackage[all]{xy}
\usepackage{amsmath}
\usepackage{amssymb}
\newcommand{\be}{\begin{equation}}
\newcommand{\ee}{\end{equation}}
\newcommand{\ben}{\begin{eqnarray}}
\newcommand{\een}{\end{eqnarray}}
\newcommand{\bes}{\begin{subequations}}
\newcommand{\ees}{\end{subequations}}

\newcommand{\bb}{\bibitem}
\begin{document}

\title{Kinklike structures in scalar field theories: from one-field to two-field models}
\author{D. Bazeia,$^{a,b,c}${{\footnote{bazeia@fisica.ufpb.br}}}  L. Losano,$^{b,c}$ and J.R.L. Santos$^{b,d}$}
\affiliation{{\small {$^a$Instituto de F\'\i sica, Universidade de S\~ao Paulo, 05314-970 S\~ao Paulo SP, Brazil\\
$^b$Departamento de F\'{\i}sica, Universidade Federal da Para\'{\i}ba, 58051-970 Jo\~ao Pessoa PB, Brazil\\
$^c$Departamento de F\'\i sica, Universidade Federal de Campina Grande, 58109-970 Campina Grande PB, Brazil\\
$^d$Department of Physics, University of Rochester, Rochester, New York 14627, USA}
}}

\date{\today}
\begin{abstract}
In this paper we study the possibility of constructing two-field models from one-field models. The idea is to start with a given one-field model and use the deformation procedure to generate another one-field model, and then couple the two one-field models nontrivially, to get to a two-field model, together with some explicit topological solutions. We show with several distinct examples that the procedure works nicely and can be used generically.
\end{abstract}


\maketitle

\section{Introduction}

Topological solutions known as kinks, vortices and monopoles are of direct interest to several areas of nonlinear science; see, e.g., \cite{b1,b2,su,W,M}. They appear in models describing spontaneous symmetry breaking, inducing phase transitions that could be used, for instance, to describe cosmic evolution in the early universe. In the simplest case of kinks, one usually requires a single real scalar field, which in the presence of spontaneous symmetry breaking can be used to mimic the Higgs field \cite{b1,b2} or to map degrees of freedom in polymers \cite{su} and in Bose-Einstein condensates \cite{M}. 

The basic model described by a real scalar field can be further extended to the case of two real scalar fields, giving rise to more sophisticated models and topological structures, again of great interest to nonlinear science. However, the two-field models are much harder to be solved, and for this reason in the current work we investigate the presence of defect structures in models described by two real scalar fields, owing to construct new models, together with the respective topological solutions. We concentrate on kinks, which are classically stable static solutions that appear when the potential is a non negative function of the scalar fields that define the model under consideration. The models that we consider admit Bogomolnyi-Prasad-Sommerfeld solutions \cite{bps}, known as BPS states, which solve first-order differential equations, leading us with bosonic portions of more sophisticated supersymmetric theories. Also, the presence of two real scalar fields makes the investigation more realistic, enhancing the power for applications in a diversity of scenarios, as one can see, e.g., in Refs.~\cite{b1,b2,W,M,bps,11,12,13,at,14,cs,15,16,J,T,17,jp,21,orbit,S,31,32} and in other works quoted therein. 

A key issue concerning the presence of defect structures in models engendering two real scalar fields is that one has to solve the equations of motion, which are two coupled second order ordinary nonlinear differential equations. To help dealing with this, the trial orbit method was proposed in \cite{12}, but there one faces an intrinsic difficulty, which concerns the presence of coupled second order differential equations. This method was later shown to be very efficient, when adapted to first order differential equations, which appear in the search of BPS states \cite{bps}, valid when the potential $V$ is non negative and can be written as the derivative of another function, which we identify
as $W$. This is explained in Ref.~\cite{orbit}, and we also quote \cite{shif} for related investigations on this issue.

Our main motivation in the present work is to use the deformation procedure introduced in \cite{deform}, taking it to construct models described by two real scalar fields, starting from a simpler model, described by a single real scalar field. As we are going to show below, it is possible to implement a general procedure, from which one starts with a single real scalar field, and use it to construct systems described by two real scalar fields. The approach relies on deforming the one-field model, to get another one-field model, and then coupling these two one-field model to end up with a two-field model, which we then solve easily. 

An important issue related to the current work is that models described by two fields are more sophisticated and can describe junctions of defects \cite{15,16,J,T,17,jp}. Also, the procedure is of direct interest to generate braneworld solutions, in a five dimensional AdS geometry with an extra dimension of infinite extent, and to produce bifurcation and pattern changing \cite{branebifur}.

For pedagogical reasons, we organize the work as follows: we start the investigation with one and two real scalar field models, briefly reviewing the BPS approach and some general aspects about the deformation procedure in Secs.~\ref{sec:2} and \ref{sec:3}, respectively. In Sec.~\ref{sec:4} we introduce the method and we study several examples in Sec.~\ref{sec:5}. We end the work in Sec.~\ref{sec:6}, where we include some comments and conclusions.

\section{Generalities}
\label{sec:2}

Let us first review some aspects relative to one and two real scalar fields in Minkowski spacetime. First, we introduce the Lagrangian density 
\be \label{lag_1}
{\cal L}=\frac{1}{2}\,\partial^\mu{\phi}\,\partial_\mu\phi-V(\phi)\,,
\ee
with $\mu=0,1$, $\partial_\mu=\partial/\partial x^\mu$, $x^\mu=(x^0=t, x^1=x)$ and $\phi=\phi(x,t)$ stands for the real scalar field. We work with dimensionless fields and coordinates. By minimizing the action, we find the equation of motion
\be
\ddot{\phi}-\phi^{\,\prime\prime}=-\frac{\partial V}{\partial \phi}\,,
\ee
where we are using the standard notation, with dots representing derivatives with respect to time and primes standing for derivatives relative to the spatial coordinate. If we work with static solutions, we are led to
\be
\phi^{\prime\prime}=\frac{\partial V}{\partial \phi}\,.
\ee

Now, we use the function $W=W(\phi)$ to write $V(\phi)$ as
\be
V(\phi)=\frac12\, {W_\phi^2}\,,
\ee
with
\be
W_\phi=\frac{d W}{d \phi}\,.
\ee
Here it is straightforward to derive that
\be\label{foe}
\phi^{\prime}=\pm\,W_\phi\,,
\ee
are first-order diferential equations which solve the equation of motion.
 
The energy density for static solution can be written in the form
\ben
\varepsilon(x)&=&\frac12\phi^{\prime\,2}+\frac12 W_\phi^2\nonumber\\ 
&=&\frac12(\phi^{\prime}\mp W_\phi)^2\pm \frac{dW}{dx}.
\een
Thus, the minimum energy configuration represents defect structure that solves the first order Eq.~\eqref{foe}
and has energy given by
\be
E_{BPS}=\left|W(\phi(\infty))-W(\phi(-\infty))\right|\,.
\ee

The same idea works for two scalar fields. In this case we introduce the model described by the two fields,  $\phi(x,t)$ and $\chi(x,t)$, in the form
\be
{\cal L}=\frac{1}{2}\,\partial^\mu{\phi}\,\partial_\mu\phi+\frac{1}{2}\,\partial^\mu{\chi}\,\partial_\mu\chi-V(\phi,\chi)\,.
\ee
We deal with static fields, and the equations of motion become
\be
\phi^{\prime\prime}=\frac{\partial V}{\partial \phi};\qquad\qquad\chi^{\prime\prime}=\frac{\partial V}{\partial \chi}\,.
\ee
We consider the potential in the form
\be \label{pot}
V(\phi,\chi)=\frac12{W_\phi^2}+\frac12{W_\chi^2}\,,
\ee
and now the first-order equations can be written in the form
\be\label{2oe}
\phi^{\prime}=\pm W_\phi;\qquad\qquad\chi^{\prime}=\pm W_\chi\,.
\ee
Here the energy density is given by
\ben
\varepsilon(x)&=&\frac{1}{2}\phi^{\prime\,2}+\frac12 \chi^{\prime\,2}+\frac12 W_\phi^2+\frac12W_\chi^2\nonumber\\
&=&\frac12 (\phi^{\prime}\mp W_\phi)^2+\frac12 (\chi^{\prime}\mp W_\chi)^2\pm \frac{dW}{dx}\,,
\een
and we see the energy is minimized for solutions to the first order Eqs.~\eqref{2oe}, attaining the value

\be
E_{BPS}=\left|W(\phi(\infty),\chi(\infty))-W(\phi(-\infty),\chi(-\infty))\right|\,.
\ee

An interesting aspect about the two field model is that we can use the integrating factor to determine an analytical orbit equation, relating the two fields $\phi(x,t)$ and $\chi(x,t)$. In order to implement it, let us work with the first order Eqs.~{\eqref{2oe}}; we use them to write
\be \label{phi_chi}
\phi_\chi=\frac{d\phi}{d\chi}=\frac{W_\phi(\phi,\chi)}{W_\chi(\phi,\chi)}\,,
\ee
This is a central point in this work, which have inspired us to propose and solve the two-field models that we investigate in Secs.~\ref{sec:4} and \ref{sec:5}.

\section{Deformation procedure}
\label{sec:3}

Let us now review the main features of the deformation procedure, as given in Ref.~\cite{deform}. We consider the model
\be
{\cal L}=\frac{1}{2}\,\partial^\mu{\phi}\,\partial_\mu\phi-V(\phi)\,,
\ee
where 
\be
V(\phi)=\frac12 W_\phi^2
\ee
and
\be \label{def_1}
\phi^{\prime}=W_\phi(\phi)\,.
\ee
We introduce another one-field model, described by
\be
{\cal L}_d=\frac{1}{2}\,\partial^\mu{\chi}\,\partial_\mu\chi-U(\chi)\,,
\ee
where
\be
U(\chi)=\frac12 {W}_\chi
\ee
and
\be \label{fochi}
\chi^{\prime}={W}_\chi(\chi)\,.
\ee
The deformation procedure requires that the two fields are related to each other through the deformation function,
that is, we suppose that there is an invertible function $f(\chi)$ such that
\be 
\phi=f(\chi)\,,
\ee 
Thus, we get
\be \label{def_2.1}
\phi^{\prime}=\frac{df}{d\chi}\,\chi^{\prime}.
\ee
For the potential $U(\chi)$ we use
\be
U(\chi)=\frac{V(\phi\rightarrow\chi)}{f_\chi^{\,2}}\,,
\ee
and now we can write 
\be \label{def_2}
W_\phi(\phi\rightarrow\chi)=W_\phi(\chi)=\frac{df}{d\chi}\,{W}_\chi(\chi)\,.
\ee

\section{The New Method}
\label{sec:4}

The procedure that we want to introduce is based in the statement that, if we use the above Eqs.~$(\ref{def_2.1})$ and $(\ref{def_2})$, we can write 
\be \label{res_1}
\frac{df}{d\chi}=\frac{\phi^{\,\prime}(\chi)}{\chi^\prime(\chi)}=\frac{d\phi}{d\chi}=\frac{W_\phi(\chi)}{{W}_\chi(\chi)}\,.
\ee
We see that this structure is similar to the one presented in Eq.~$(\ref{phi_chi})$, for the two-field model. Thus, we get inspiration on this to include the key idea of our method, which relies on the use of the deformation function in order to rewrite $(\ref{res_1})$ as
\be     
\frac{d\phi}{d\chi}=\frac{{W}_\phi(\phi,\chi)}{{W}_\chi(\phi,\chi)}\,,
\ee
which would give us an orbit relation for the two-field model which we are proposing. To make this idea to work, we first recognize that the first order differential equation $(\ref{def_1})$ can be written in one of the three distinct but equivalent ways 
\be\label{aa}
\phi^{\prime}=W_\phi(\phi)\,, \;\;\; \phi^\prime=W_\phi(\chi)\,,\;\;\; \phi^{\prime}=W_\phi(\phi,\chi)\,,
\ee
where in the second expression we have changed $\phi\to f(\chi)$ everywhere, to make $W_\phi$ a function of $\chi$ alone, and in the third expression we have changed $\phi\to f(\chi)$ partially, that is, we have changed the field $\phi$ which appear in $W_\phi(\phi)$ in a particular way, making $W_\phi$ a specific function of the two fields $\phi$ and $\chi$, coupling the two fields. This is the key step of the method, and we illustrate the issue as follows: if $W_\phi(\phi)$ contains the term $\phi^3$, for instance, we can write $\phi^3=\phi\times\phi^2$, and we can change this as $\phi\times f^2(\chi)$ or $\phi^2\times f(\chi)$, introducing distinct couplings between the two fields, leading to distinct models. The same procedure can be used for $(\ref{fochi})$, and we get
\be\label{bb}
\chi^{\prime}=W_\chi(\chi)\,,\;\;\; \chi^{\prime}=W_\chi(\phi)\,, \;\;\; \chi^{\prime}=W_\chi(\phi,\chi)\,.
\ee

Since the third step in the above two expressions \eqref{aa} and \eqref{bb} can be implemented at will, we now work to construct a mechanism to control the procedure as follows: we introduce three sets of three real parameters, ${a_1,a_2,a_3}$,
${b_1,b_2,b_3}$, and ${c_1,c_2,c_3}$, such that $a_1+a_2+a_3=1$, $ b_1+b_2+b_3=1$, and $c_1+c_2+c_3=0$. We then make the changes $W_\phi\to a_1\,W_\phi(\chi)+a_2\,W_\phi(\phi,\chi)+a_3\,W_\phi(\phi)$ and $W_\chi\to b_1\,W_\chi(\chi)+b_2\,W_\chi(\phi,\chi)+b_3\,W_\chi(\phi)$, and we write

\begin{widetext}
\be\label{w}
\frac{d\phi}{d\chi}=\frac{W_\phi}{W_\chi}=\frac{a_1\,W_\phi(\chi)+a_2\,W_\phi(\phi,\chi)+a_3\,W_\phi(\phi)+c_1\,g(\chi)+c_2\,g(\phi,\chi)+c_3\,g(\phi)}{b_1\,W_\chi(\chi)+b_2\,W_\chi(\phi,\chi)+b_3\,W_\chi(\phi)}\,,
\ee

\bigskip

\end{widetext}
\noindent
where $g(\phi)=g(\chi)=g(\phi,\chi)$ is in principle an arbitrary function, constructed in the same way we did to write the three expressions for $W_\phi$ and $W_\chi$.
Instead of adding the term $c_1\,g(\chi)+c_2\,g(\phi,\chi)+c_3\,g(\phi)$ to the numerator of \eqref{w}, we could add it to the denominator, but this would only change the role between the two fields $\phi$ and $\chi$. The specific form of $g$ will be obtained from the constraint to be given below, obtained from the requirement that the potential of the two-field model is described by the function $W(\phi,\chi)$ which obeys
\be\label{wc}
W_{\phi\chi}=W_{\chi\phi}
\ee

Since we are searching for two-field models, the two fields must couple with each other, so we have to write $W_\phi(\phi,\chi)$ and $W_\chi(\phi,\chi)$ in the form of products involving the two fields $\phi$ and $\chi$. 

We see from the above expression $\eqref{w}$ that we are changing $W_\chi$ for
\be \label{wchi}
b_1\,W_\chi(\chi)+b_2\,W_\chi(\phi,\chi)+b_3\,W_\chi(\phi)\,.
\ee
Also, we are changing $W_\phi$ for
\ben \label{wphi}
a_1W_\phi(\chi)&\!+\!&a_2W_\phi(\phi,\chi) +a_3W_\phi(\phi)\nonumber\\
&+&c_1g(\chi)+c_2g(\phi,\chi)+c_3g(\phi)\,. 
\een
However, we have to impose $\eqref{wc}$, which leads us with the constraint
\ben\label{cons}
&&\!\!\!b_2\,W_{\chi\phi}(\phi,\chi)+b_3\,W_{\chi\phi}(\phi)=\nonumber\\
&&\!\!\!a_1\,W_{\phi\chi}(\chi)\!+\!a_2\,W_{\phi\chi}(\phi,\chi)\!+\!c_1\,g_\chi(\chi)\!+\!c_2\,g_\chi(\phi,\chi)\,\;\;\;\;
\een
which is used to calculate the function $g$, since we already know both $W_\phi$ and $W_\chi$.
The procedure allows us to determine the final form for $W(\phi,\chi)$, to define the proposed two-field model, together with the corresponding defect structure
it comprises, by construction. This ends the procedure, so we focus on some examples in the next section.

\section{Examples}
\label{sec:5}

To see how the method works, let us now illustrate the procedure with several examples, which we describe below.
\subsection{Example 1: $\phi^4$ versus $\phi^4$} 
\label{sec:51} 

The idea here is to construct one two-field model from two one-field models, having fourth-order power in each field. We start considering the one-field model, described by the real scalar field $\phi$, with $W$ such that
\be
\phi^\prime=W_\phi=a\,(1-\phi^2)\,,
\ee
which gives the kinklike solution
\be
\phi(x)=\tanh(a\,x)\,.
\ee
Here $a$ is a real parameter, dimensionless. This is the standard $\phi^4$ model, with spontaneous symmetry breaking and we are using dimensionless units.

Now, let us deform this model to get to another one-field model. We consider the deformation function that follows
\be
\phi=f(\chi)=\sqrt{1-\frac{\chi^2}{b^2}}\,.
\ee
where $b$ is another parameter, which controls the deformation function. This leads us to the first-order equation
\be
\chi^{\prime}=W_\chi=- a\,\chi\,\sqrt{1- \frac{\chi^2}{b^2}}\,.
\ee
The solution is now given by
\be
\chi(x)=b\, \text{sech} (a\,x)\,.
\ee

The next step is to write the three distinct forms of the first-order differential equations, for both $\phi$ and $\chi$. We use the deformation function to write
\bes\ben
W_\phi(\phi)&=&a\,\left(1-\phi ^2\right)\\
W_\phi(\chi)&=&\frac{a}{b^2}\,{\chi^2}\,,\\
W_\phi(\phi,\chi)&=&\frac{a}{b}\,{\chi}\,\sqrt{1-\phi^2}\,,
\een\ees
as well as,
\bes\ben
W_\chi(\chi)&=&- a\,\chi\,  \sqrt{1-\frac{\chi ^2}{b^2}}\,,\\
W_\chi(\phi)&=&-a\,b\,\phi\,\sqrt{1-\phi^2}\,,\\
W_\chi(\phi,\chi)&=&-a\,\chi\,  \phi\,.
\een\ees

If we want to avoid the presence of the square root in the final expression of the potential, we consider $a_2=b_1=b_3=0$. Also, we take $c_2=0$ in $(\ref{wphi})$, and so we have $a_1+a_3=1$, $b_2=1$, and $c_1=-c_3$. Therefore, by using the constraint  $(\ref{cons})$ we determine that
\be
g(\chi)=-\frac12\,\frac{a}{c_1}\left(1+2\,\frac{a_1}{b^2}\right)\chi^2,
\ee
and the deformation function allows us to obtain
\be
g(\phi)=-\frac12\,\frac{a\,b^2}{\,c_1}\left(1+2\,\frac{a_1}{b^2}\right)(1-\phi^2)\,.
\ee
Putting this results back into $(\ref{wphi})$, we find
\be
W_\phi=-\frac{a}{2}\,\chi^2+a\,\left(1+\frac{b^2}{2}\right)\,(1-\phi^2),
\ee
and from $(\ref{wchi})$, we have
\be
W_\chi=-a\,\chi\,\phi\,.
\ee

Thus, we can perform simple integrations to determine the final form of our two scalar fields superpotential, which is 
\be \label{wrmap}
W(\phi,\chi)= a\,\left(1+\frac12{b^2}\right)\,\left(\phi-\frac13{\phi^3}\right)-\frac12{a}\,\phi\chi^2\,.
\ee
This is the function which defines the two-field model. And more, the model has the static solution
\be
\phi(x)=\tanh(a\,x)\,, \qquad\chi(x)=b\, \text{sech} (a\,x)\,.
\ee
We see that if we make the identification
\be
a=2\,r \qquad \text{and} \qquad b=\pm\sqrt{\frac{1}{r}-2}\,
\ee
with $r\in(0,1/2)$ we get
\be \label{bnrt}
W_r(\phi,\chi)=\phi-\frac13\phi^3-r\,\phi\,\chi^2\,,
\ee
and the solutions
\bes\ben \label{orb_bnrt}
\phi(x)&=&\tanh(2\,r\,x),\\
 \chi(x)&=&\pm\,\sqrt{\frac{1}{r}-2}\, \,\text{sech} (2\,r\,x)\,.
\een\ees
This model was investigated before and used in several distinct applications; see, e.g., Refs.~\cite{14,17,21}.

\subsection{Example 2: $\phi^4$ versus $\chi^6$}
\label{sec:52}
The next example is constructed through a combination between $\phi^4$ and $\chi^6$ models. Here, we start with
\be
\phi^\prime=W_\phi=a^2-(\phi-a)^2\,,
\ee
which gives the defect structure
\be
\phi(x)=a+a\,\tanh(a\,x)\,,
\ee

Moreover, we consider the deformation function 
\be
\phi=f(\chi)=2\,a-\frac{a}{b^2}\,\chi^2\,,
\ee
thus, by applying the deformation method we obtain the first order differential equation
\be
\chi^{\prime}=W_\chi=-\frac{a}{2}\,\chi\,\left(2-\frac{\chi^2}{b^2}\right)\,,
\ee
with the topological solution
\be
\chi(x)=b\,\sqrt{1-\tanh(a\,x)}\,.
\ee

The procedure requires that we write 
\begin{widetext}

\be
W_\phi(\phi)=a^2-(\phi-a)^2\,,\qquad
W_\phi(\chi)=\frac{a^2}{b^2}\,\left(2\,\chi^2-\frac{\chi^4}{b^2}\right)\,,\qquad
W_\phi(\phi,\chi)=\frac{a^2\,\chi^2}{b^2}\left(\frac{\chi^2}{b^2}+2\,\frac{\phi-a}{a}\right)\,,
\ee
and
\be
W_\chi(\chi)=-\frac{a}{2}\,\left(2-\frac{\chi^2}{b^2}\right)\,\chi\,, \qquad
W_\chi(\phi,\chi)=-\frac{a}{2}\left(1+\frac{\phi-a}{a}\right)\,\chi\,.
\ee
Here we used $b_3=0$, since we want to avoid the square root in the two-field model, then $b_1+b_2=1$.
We also choose $c_1=0$, so we have $c_3=-c_2$; the constraint \eqref{cons} then gives
\be
g(\phi,\chi)=-\frac{b_2}{4\,c_2}\,\chi^2-\frac{a_2\,a^2}{c_2}\,\frac{\chi^2}{b^2}\,\left(\frac{\chi^2}{b^2}+2\,\frac{\phi-a}{a}\right)-\frac{a_1}{c_2}\frac{a^2}{b^2}\,\left(2\,\chi^2-\frac{\chi^4}{b^2}\right)\,,
\ee
and we can use the deformation function to rewrite $g(\phi,\chi)$ as follows 
\be
g(\phi)=-\frac{b_2\,b^2}{4\,c_2}\,\left(1-\frac{\phi-a}{a}\right)
-\frac{a^2\,(a_2+a_1)}{c_2}\,\left(1-\frac{(\phi-a)^2}{a^2}\right)\,. 
\ee

With the above result, we then have all the ingredients to determine $W(\phi,\chi)$. After some calculations we get
\be
W(\phi,\chi)=-\frac{(1-b_2)\,a}{2}\,\left(\chi^2-\frac{\chi^4}{4\,b^2}\right)
-b_2\,{\phi}\,\frac{\chi^2}{4}\,+\frac{b_2\,b^2}{4}\,\left(2\,\phi-\frac{\phi^2}{2\,a}\right)
+\,\left(a\,\phi^2-\frac{\phi^3}{3}\right)\,.
\ee

\end{widetext}

Several  interesting models can be determined by taking different values for $a$, $b$, and $b_2$. In particular, if we choose $b^2={2}$, $a=\pm\,1/2$ and $b_2=1$, we get
\be \label{bnrt2}
W(\phi,\chi)=\phi-\frac{\phi^3}{3}-\frac{\chi^2}{4}\,\phi\,,
\ee
which is the previous model, for $r=1/4$; see $(\ref{bnrt})$. Here, however, we have the solutions
\be \label{orb_3} 
\phi(x)=-\frac{1}{2}+\frac12\,\tanh\left(\frac{x}{2}\right),\, \chi(x)=\pm\sqrt{2\,+2\,\tanh\left(\frac{x}{2}\right)}
\ee
and
\be \label{orb_4}
\phi(x)=\frac{1}{2}+\frac12\,\tanh\left(\frac{x}{2}\right)\,,\,\, \chi(x)=\pm\sqrt{2\,-2\,\tanh\left(\frac{x}{2}\right)}\,.
\ee

\subsection{Example 3: $\phi^4$ versus $\chi^3$}

In this example we explore models having third and fourth power in the fields. We start with
\be
\phi^{\prime}=W_\phi=1-\phi^2\,.
\ee
The solution is
\be
\phi(x)=\tanh(x)\,.
\ee
We consider the deformation function,
\be\label{def3}
\phi=\sqrt{1-\frac{\chi}{a}}\,,
\ee
and we obtain
\be
\chi^{\prime}=W_\chi=-2\,\chi\,\sqrt{1-\frac{\chi}{a}\,}\,,
\ee
which is solved by
\be
\chi(x)=a\,\text{sech}^2(x)\,.
\ee

Using the orbit \eqref{def3}, we obtain the equations
\be
W_\phi(\phi)=1-\phi^2\,,\,\,\,  W_\phi(\chi)=\frac{\chi}{a}\,,
\ee
and
\be
W_\chi(\phi,\chi)=-2\,\phi\,\chi\,,\,\,\, W_\chi(\phi)=-2a\,\phi(1-\phi^2)\,, 
\ee
since we are avoiding the presence of the square root in the two-field model.
These choices lead to $a_2=b_1=c_1=0$, then $a_3+a_1=1$,  $b_2+b_3=1$, and $c_3=-c_2$. Thus, we can write 
the function $g(\phi,\chi)$ as
\be
g(\phi,\chi)=-\frac{2ab_3}{c_2}(1-3\phi^2)\chi-\frac{b_2}{c_2}\chi^2-\frac{a_1}{ac_2}\,\chi\,.
\ee
We can use the deformation function to rewrite it in terms of the $\phi$ field alone, in the form
\ben
g(\phi)&=&\left(-\frac{2a^2b_3}{c_2}(1-3\phi^2)-\frac{a^2b_2}{c_2}(1-\phi^2)-\frac{a_1}{c_2}\right)\nonumber\\
&&\times\, (1-\phi^2)\,.
\een
With these results we find
\ben
W(\phi,\chi)&=&(1+2a^2-a^2b_2)\phi+(1+8a^2-6a^2b_2)\frac{\phi^3}{3}\nonumber\\
&&-2a(1-b_2)(1-3\phi^2)\phi\chi-b_2\phi\chi^2\nonumber\\
&&- (6-5b_2)a^2\frac{\phi^5}{5}\,,
\een
which leads to the expressions
\ben
W_\phi&=&(1-\phi^2)\left(1+(2-b_2)a^2-(6-5b_2)a^2\phi^2\right)\nonumber\\
&&-2a(1-b_2)(1-3\phi^2)\chi-b_2\chi^2,
\een
and
\be
W_\chi= -2a(1-b_2)(1-3\phi^2)\phi-2b_2\phi\chi.
\ee
These results allow us to calculate the potential $V(\phi,\chi)$, as dictated by Eq.~$(\ref{pot})$.  
\\
\subsection{Example 4: p-model}

Our final example describes a generalization of the p-model, as introduced in \cite{pmodel}. Here, we start with
\be
\phi^{\,\prime}=W_\phi=p\,(\phi^{(p-1)/p}-\phi^{(p+1)/p})\,
\ee
where $p=1,3,5,...$ is odd integer. Note that for $p=1$ we get back to the standard $\phi^4$ model.
In general, however, we have an interesting model, and we have the 2-kink solution
\be \label{solp1}
\phi(x)=\tanh^p(x)\,,
\ee
as found in \cite{pmodel}. This model is more complicated then the previous models, so we perform the simpler deformation
\be
\phi=f(\chi)=\frac{\chi}{a}\,,
\ee
which leads us to
\be
\chi^{\prime}=W_\chi=p\,a\,[(\chi/a)^{(p-1)/p}-(\chi/a)^{(p+1)/p}]\,,
\ee
with analytical solution given by
\be \label{solp_2}
\chi(x)=a\,\tanh^p(x)\,.
\ee

The next step is to write the first-order equations; they are constructed with the distinct functions
\begin{widetext}
\be
W_\phi(\phi)=p\,(\phi^{(p-1)/p}-\phi^{(p+1)/p}),\,W_\phi(\chi)=p\,\left[\left(\frac{\chi}{a}\right)^{(p-1)/p}-\left(\frac{\chi}{a}\right)^{(p+1)/p}\right],\,
W_\phi(\phi,\chi)=p\,\left[\left(\frac{\chi}{a}\right)^{(p-1)/p}-\phi\,\left(\frac{\chi}{a}\right)^{1/p}\right],
\ee
and
\be
W_\chi(\chi)=pa\left[\left(\frac{\chi}{a}\right)^{(p-1)/p}-\left(\frac{\chi}{a}\right)^{(p+1)/p}\right],\,
W_\chi(\phi)=pa(\phi^{(p-1)/p}-\phi^{(p+1)/p}),\,
W_\chi(\phi,\chi)=pa\left[\left(\frac{\chi}{a}\right)^{(p-1)/p}-\phi\left(\frac{\chi}{a}\right)^{1/p}\right].
\ee
Therefore, if we consider the constraint with $c_1=0$ and $b_3=0$ to avoid  negative exponent in the potential, we set $c_3=-c_2=0$ and $b_1+b_2=1$, in order to obtain
\be
g(\phi,\chi)=-\frac{b_2}{c_2}\frac{\,p^2\,a^2}{p+1}\left(\frac{\chi}{a}\right)^{{(p+1)/p}} -\frac{a_2}{c_2}\,p\,\left[\left(\frac{\chi}{a}\right)^{{(p-1)/p}}-\phi\,\left(\frac{\chi}{a}\right)^{1/p}\right]-\frac{a_1}{c_2}\,p\,\left[\left(\frac{\chi}{a}\right)^{{(p-1)/p}}-\left(\frac{\chi}{a}\right)^{{(p+1)/p}}\right]\,.
\ee
As before, we can use the deformation function to write
\be
g(\phi)=-\frac{b_2}{c_2}\frac{\,p^2\,a^2}{p+1}\,\phi^{{(p+1)/p}}-\frac{a_2+a_1}{c_2}\,p\,(\phi^{(p-1)/p}-\phi^{(p+1)/p})\,. 
\ee

We follow the above procedure to obtain
\ben
W(\phi,\chi) &=& b_1\,p^2\,a^2\,\left[\frac{1}{2\,p-1}\,\left(\frac{\chi}{a}\right)^{(2\,p-1)/p}-\frac{1}{2\,p+1}\,\left(\frac{\chi}{a}\right)^{(2\,p+1)/p}\right]+b_2\,p^3\,\frac{a^2\,\phi^{(2\,p+1)/p}}{(p+1)\,(2\,p+1)}\\ \nonumber
&&
+b_2\,p^2\,a^2\,\left[\frac{1}{2\,p-1}\,\left(\frac{\chi}{a}\right)^{(2\,p-1)/p}-\frac{\phi}{p+1}\,\left(\frac{\chi}{a}\right)^{(p+1)/p}\right]+p^2\,\left[\frac{\phi^{(2\,p-1)/p}}{2\,p-1}-\frac{\phi^{(2\,p+1)/p}}{2\,p+1}\right]\,.
\een

These results allow us to construct the pair
\be
W_\phi=p\,\left(\phi^{(p-1)/p}-\phi^{(p+1)/p}\right)+\frac{b_2\,p^2\,a ^2}{p+1}\left[\phi ^{(p+1)/p } 
-\left(\frac{\chi }{a }\right)^{(p+1)/p}\right]\,,
\ee
and
\be
W_\chi= b_1\,p\,a\left[\left(\frac{\chi}{a}\right)^{(p-1)/p}-\left(\frac{\chi}{a}\right)^{(p+1)/p}\right]+b_2\,p\,a\,\left[\left(\frac{\chi}{a}\right)^{(p-1)/p}-\phi\,\left(\frac{\chi}{a}\right)^{1/p}\right]\,.
\ee
\end{widetext}
Consequently, we are able to determine the potential $V(\phi,\chi)$ and construct the corresponding two-field model. It is interesting to note that if we take $p=3$, $b_1=0$, $b_2=1$ and $a=1$ in $W_\phi$ and $W_\chi$, we get to
\be 
W_\phi=3\,\phi^{2/3}-\frac{3}{4}\,\phi^{4/3}-\frac{9}{4}\,\chi^{4/3}\,; \qquad W_\chi = 3\,\chi^{2/3}-3\,\phi\,\chi^{1/3}
\ee
and so we get
\be
W(\phi,\chi)=\frac{9}{5}\,\left(\phi^{5/3}+\chi^{5/3}\right)-\frac{9}{28}\,\phi^{7/3}-\frac{9}{4}\,\phi\,\chi^{4/3}\,.
\ee
The solutions in this case are
\be
\phi(x)=\tanh^3(x)\qquad \text{and} \qquad\chi(x)=\tanh^3(x)\,.
\ee
This example shows for the first time an interesting model where the topological solution appears as a coupling of two 2-kink structures.
Evidently, we can obtain many other new models for distinct values of $p$ and the other parameters.

\section{Final Comments}
\label{sec:6}

In this work we proposed a new procedure to generate two-field models. The method starts with a given one-field model, which is used to generate another one-field model, via the deformation procedure introduced in Ref.~\cite{deform}. We then couple the two one-field model to generate a two-field model. The procedure is illustrated with several distinct examples, to show how efficient the method is, to construct new two-field models. An important advantage of the procedure is that it automatically gives some analytical solutions for these new systems.

The current investigation poses some interesting issues, one of them concerning extensions of the method to construct models described by three or more real scalar fields, and models described by non polynomial potentials. Another issue is related to cosmology, and the two-field models can be used to model interactions between dark matter and dark energy, as investigated for instance in Ref.~\cite{dmde}. Some of these issues are now under consideration, and we hope to report on them in the near future.

\acknowledgments

The authors would like to thank A. de Souza Dutra for discussions, and CAPES, CNPq and FAPESP for partial financial support.


\end{document}